\documentclass[
preprint,
aps,
showpacs,
nofootinbib,
showkeys,
byrevtex,
letterpaper]{revtex4}

\usepackage{amsmath}
\usepackage{epsfig}
\usepackage{latexsym}

\def\){\right)} 
\def\({\left(} 
\def\]{\right]} 
\def\[{\left[}

\def\CPT{$\chi${\rm PT}}
\def\CPTs{$\chi${\rm PT }}
\def\CSW{c_{\rm SW}}
\def\psibar{\overline{\psi}}

\newcommand{\eqn}[1]{\label{eq:#1}}
\newcommand{\refeq}[1]{(\ref{eq:#1})} 
\newcommand{\eq}{eq.~\refeq}
\newcommand{\eqs}[2]{eqs.~(\ref{eq:#1}-\ref{eq:#2})}
\newcommand{\eqsii}[2]{eqs.~(\ref{eq:#1}),~(\ref{eq:#2})}

\newcommand{\Eq}{Eq.~\refeq} 

\newcommand{\beq}{\begin{eqnarray}}
\newcommand{\eeq}{\end{eqnarray}}

\newcommand{\mcal}[1]{{\mathcal #1}}

\newcommand{\makefig}[4]{\begin{figure}[t] 
                           \centerline{\epsfysize=#3 in \epsfbox{#2}} 
                           \caption{#4 } 
                           \label{#1}
                         \end{figure}}

\begin{document}
\preprint{MIT-CTP 3319}
\preprint{LBNL-51684}
\preprint{BUHEP-02-36}
\title{Simulations with different lattice Dirac operators for valence and sea quarks
}
\author{Oliver B\"ar~${}^{a}$\footnote{Email: {\tt obaer@lns.mit.edu}},  
Gautam Rupak~${}^{b}$\footnote{Email: {\tt grupak@lbl.gov}} 
and
Noam Shoresh~${}^{c}$\footnote{Email: {\tt shoresh@bu.edu}}
}
\affiliation{${}^a$ Center for Theoretical Physics, Massachusetts Institute of
Technology, Cambridge, MA 02139\\
${}^b$Lawrence Berkeley National Laboratory, Berkeley, CA 94720 \\
${}^c$Department of Physics, Boston University, Boston, MA 02215}

\begin{abstract}

We discuss simulations with different lattice Dirac operators for sea
and valence quarks.   
A goal of such a ``mixed'' action approach is to  
probe deeper the chiral regime of QCD 
by enabling simulations with light
valence quarks.
This is achieved by using chiral fermions as valence quarks while
computationally inexpensive fermions are used in the sea sector.
Specifically, we consider Wilson sea quarks and
Ginsparg-Wilson valence quarks. 
The local Symanzik action for this mixed theory is derived to
$\mcal{O}(a)$, and
the appropriate low energy chiral effective Lagrangian
is constructed, including the leading $\mcal O(a)$ contributions. 
Using this Lagrangian one can calculate expressions for physical
observables and determine the Gasser-Leutwyler coefficients by
fitting them to the lattice data.

\end{abstract}

\maketitle 

\begin{section}{Introduction}
\label{introduction}

In order to extract predictions of QCD from numerical methods with
controlled systematic errors, a lattice formulation is
required for which the sources of deviations from QCD are understood and are under
control. 
A significant source of systematic errors for present day lattice simulations
are the light quark masses. Even the most powerful computers
today do not allow simulations with up- and down-quark masses as light as
realized in nature.
Instead one simulates with heavier quark masses and fits the
analytic predictions obtained in chiral perturbation theory (\CPT) to
the data. The free parameters in the fit are the low
energy constants of \CPTs \cite{Gasser:1985gg}, and once
they are determined an extrapolation to small quark masses is possible
\cite{Sharpe:2000bc,Cohen:1999kk} .
Still,  to perform the chiral extrapolation the quark masses must be
small enough so that \CPTs is applicable.
In practice one would require that next-to-leading order (NLO) \CPTs
describe the data reasonably well.%
\footnote{At next-to-next-to-leading order  (NNLO) many new
  unknown parameters enter the chiral Lagrangian, which greatly
  reduces the predictive power of \CPT.}

The present lattice simulations do not meet this
requirement \cite{Bernard:2002yk,Durr:2002fe,Farchioni:2002ca}.  
The data do not show 
the characteristic curvature predicted by NLO \CPT. 
In fact the data show a rather linear behavior which either means
that higher orders 
in the chiral expansion 
are not negligible  
or worse, one is not in the chiral regime at all
(see Claude Bernard's part in \cite{Bernard:2002yk}).  
In either case, simulations with lighter quark
masses are required in order to apply \CPTs with confidence.   

Lattice simulations with light fermions, especially sea quarks, are
computationally demanding and 
the numerical cost increases substantially with decreasing quark
masses.
Realistically only the least expensive fermions, Wilson and
Kogut-Susskind, can be used on sufficiently large and fine lattices.
Lattice fermions with better chiral properties are still too expensive
to be used as sea quarks, and this situation is not likely to change
in the near future. 
It is nevertheless expected that the next generation of TFLOP
machines will make it possible to generate a few sets of unquenched
configurations with sea quarks light enough to be in the chiral
regime. 

To obtain more information from these configurations
they should (and will) be analyzed with various different valence
quark masses, i.e.~by studying PQ QCD. By
including lattice measurements with lighter valence
quarks it is possible to penetrate further the chiral regime of QCD. 
This leads to more data points and
would  allow more  reliable fits of 
PQ \CPT\cite{BGPQ}
to the lattice
data \cite{Sharpe:2000bc}.  
The reach of such simulations, however, is limited.
The cost of light valence quarks also increases with the decreasing
mass, and can become prohibitively high for quark masses that are
still not very small. 
This is particularly true for Wilson fermions 
because of the explicit chiral symmetry breaking by the Wilson term.

An interesting idea for probing the chiral regime 
is to use different
lattice fermions for the valence and sea quarks. 
In particular, by
choosing lattice fermions with good chiral properties for the valence
quarks, the valence quark mass 
 can be made much smaller than
in ordinary PQ simulations. 
A central  goal of this
strategy
is the same as of PQ QCD - 
to explore a  larger 
portion 
of the chiral regime by extracting
more data points
from a given set of  unquenched configurations (see fig.~\ref{plot}).
This should result in more reliable estimates for the low-energy
constants of \CPTs 
at NLO, the Gasser-Leutwyler coefficients.
Furthermore, one might expect to reduce the size of explicit chiral symmetry
breaking by using Ginsparg-Wilson fermions at least for the valence
quarks.
This is a computationally affordable compromise
of
the lattice
theorist's ideal of using Ginsparg-Wilson fermions for both valence and
sea quarks.
\makefig{plot}{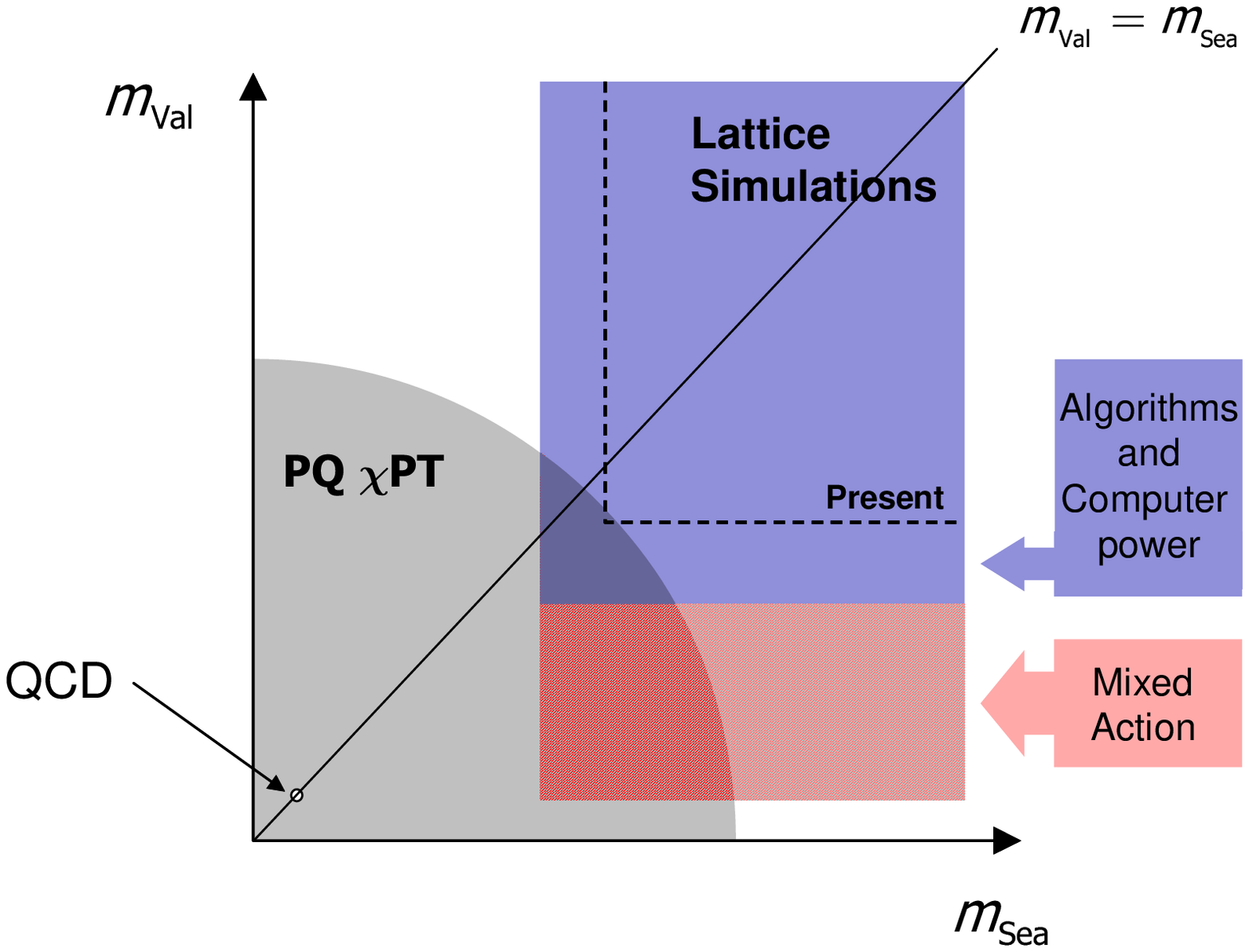}{3}{\it Qualitative representation of the
 space of quark masses. The ``chiral regime'', where 
PQ \CPTs can be applied, is the quarter-circular region. 
The upper right rectangle, limited by the dashed line, describes the
part of the space covered by present simulations. As current data
suggest, there is very little or no overlap between 
that rectangle and the chiral regime.
It is expected that improvement in algorithms and
computer power will allow reducing the sea and valence quark masses in
PQ simulations, as is represented by the enlargement of the
previous rectangle. It is possible that the chiral region will be
penetrated by such simulations, as shown by the small section of
overlap between the 
enlarged rectangle and the chiral region. Finally, using chiral
valence fermions in a mixed action simulation would make it possible to extend
the reach of simulations significantly in the direction of
lighter valence quarks.
}
 
In this paper we construct the low-energy chiral effective theory for a
``mixed'' lattice action, with explicit dependence on powers of the lattice
spacing $a$ by first constructing the appropriate local Symanzik
action. 
There are several reasons for taking this
approach. First, the defining non-orthodox feature
of the mixed action approach -- the use of different Dirac operators for the
sea and valence sectors -- is purely a lattice artifact. 
This is a consequence of the fact that
by construction all proper lattice fermions reproduce the
same continuum physics, and therefore all mixed lattice theories reduce to PQ
QCD in the  continuum  limit.
An expansion in $a$ is thus a natural tool to investigate potential peculiarities
of the mixed action formulation. Second, a theoretical understanding
of the $a$-dependence in lattice simulations can guide the continuum 
limit, or allow the extraction of physical information directly from the
lattice data, without taking the continuum limit first. Third, \CPTs
provides a useful framework for studying 
the chiral symmetry breaking due to the discrete space-time lattice.  
Effective theories of this type have been studied in several
similar contexts \cite{Sharpe:1998xm,Rupak:2002sm,Lee:1999zx,Aubin:2002ss}.

What is dubbed  here "mixed action methods'' refers to  a class of lattice
theories corresponding to  different choices of Dirac operators for
the valence  and sea 
quarks. In the next section we use a fairly simple example to illustrate the general framework of mixed lattice theories. We consider Wilson fermions for the sea quarks,  together with valence
fermions that satisfy the Ginsparg-Wilson relation.  
To describe the lattice action close to the continuum limit
we construct the local Symanzik effective action up to $\mcal{O}(a)$.
The usual arguments used in the formulation of \CPTs are then
applied to this effective action. This leads to a chiral
expansion in which the dependence on the lattice spacing is explicit.

\end{section}
\begin{section}{The Chiral effective action}\label{theory}
\subsection{Lattice action}\label{action}
In the following we always consider a hyper-cubic Euclidean space-time lattice
with lattice spacing $a$. We assume
either an infinite lattice or a finite lattice large enough that one can safely
ignore finite volume effects. 
 
The mixed lattice action that describes $N_f$ Wilson sea  and
$N_V$ Ginsparg-Wilson valence fermions has the structure
\begin{align}
S_{M}\,  =\, S_{\text{YM}} [U] + S_{W} [\psibar _{S} ,\psi_{S},U] + S_{\text{GW}}
[\psibar_{V},\psi _{V} ,U].\eqn{SH} 
\end{align}
$U$ denotes  the gauge field defined on the links of the lattice,
 $\psi_{S}$ ($\psibar _{S}$) are the sea quark (anti--quark) fields and
$\psi_{V}$ ($\psibar _{V}$) denote vectors with $N_V$ anti--commuting valence 
quarks (anti--quarks) and  $N_V$ c-number-valued ghost quarks
 (anti-quarks). 

The precise choice for the gauge field action $S_{\text{YM}}$ is irrelevant in the
following, so we leave it unspecified. For the sea quarks we
choose the Wilson action \cite{Wilson:1974sk}, given by 
\begin{align}
& S_{W}  = a^4 \sum\limits_x {\psibar _{S} (D_W  + m_{\text{Sea}} )\psi _{S} 
(x)}\,, \\ 
& D_W  = \frac{1}{2} \left\{\gamma _\mu  (\nabla _\mu^*+\nabla _\mu)  
- a\,r \nabla _\mu^*\nabla _\mu)\right\}\,,\eqn{SW}
\end{align}
where $m_{\text{Sea}}$ denotes the $N_f\times N_f$ quark  mass matrix in the sea
sector and $r$  the Wilson parameter. $\nabla _\mu^*,\,\nabla _\mu$ are the
usual covariant, nearest neighbor  backward and forward difference operators. 

The action for the valence and ghost quarks is given by
\begin{align}\eqn{SGW}
S_{GW}  = a^4\sum\limits_x {\psibar _{V} \left\{ {D_{GW}  
+ m_{\text{Val}}\,\left( {1 - \tfrac{1}
{2}aD_{GW} } \right) } \right\}\psi _{V} (x)}\,.
\end{align}
The valence and ghost quark masses are contained in the $2N_V\times 2N_V$ mass
matrix $m_\text{Val}$ of the form
$m_\text{Val}=\text{diag}(M_\text{Val},M_\text{Val})$ where
$M_\text{Val}$ is an $N_V\times N_V$ matrix (i.e. each valence quark has a
corresponding ghost field with the same mass).
The Dirac operator $D_{GW}$ is assumed to be a local operator satisfying the 
Ginsparg--Wilson relation~\cite{Ginsparg:1982bj}
\beq\label{GW_relation}
\gamma _5 D_{GW}  + D_{GW} \gamma _5  = aD_{GW} \gamma _5 D_{GW} .
\eeq
Both the fixed-point Dirac operator and the overlap operator 
satisfy this relation
\cite{Hasenfratz:1998ri,Neuberger:1998fp,Neuberger:1998wv}. For the
following discussion, however, there is no need to specify $D_{GW}$
any further.  

\subsection{Flavor symmetry of the lattice action}\label{symmetries}

When $m_\text{Sea}=0$, $m_\text{Val}=0$, and $r=0$, the flavor
symmetry group of $S_M$ is
\beq\eqn{Sym_Group_Hybrid}
SU(N_f )_L  \otimes SU(N_f )_R  \otimes SU(N_V |N_V )_L  \otimes SU(N_V |N_V )_R. 
\eeq
To see this it is convenient to write $S_{GW}$ and $S_{W}$ in terms of
chiral components. 
The right- and left-handed 
sea
quark fields are defined with the usual projectors $\tfrac{1}{2}(1\pm \gamma_5)$. 
For the valence and ghost fields, one first defines the hermitian operator  
\begin{align}
\hat{\gamma}_5\, = \, \gamma_5(1-aD_{GW})\,,\eqn{gammahat}
\end{align}
which is unitary as a consequence of (\ref{GW_relation}). 
Valence right- and left- handed fields are now defined by
\cite{Niedermayer:1998bi}   
\begin{align}
&\psibar_{V,R}\,=\,\psibar_V\frac{1}{2}(1-\gamma_5)\,,\qquad \psi_{V,R}
\,=\,\frac{1}{2}(1+\hat{\gamma_5})\psi_V\,,\\ 
& \psibar_{V,L}\,=\,\psibar_V\frac{1}{2}(1+\gamma_5)\,,\qquad \psi_{V,L}
\,=\,\frac{1}{2}(1-\hat{\gamma_5})\psi_V\,. 
\end{align}
The fermionic actions can now be rewritten as
\begin{align}\eqn{SWchiral}
S_W  = & \, a^4 \sum_x {\psibar_{S,L} \tfrac{1}
{2}\gamma _\mu  \left( {\nabla _\mu   + \nabla _\mu ^* } \right)\psi _{S,L}(x)  + \psibar_{S,R} \tfrac{1}
{2}\gamma _\mu  \left( {\nabla _\mu   + \nabla _\mu ^* } \right)\psi _{S,R}(x) }  \\
 & \hspace{.5in}+ \psibar_{S,L} \left( {m_{\text{Sea}}  - \tfrac{1}
{2}ar\nabla _\mu  \nabla _\mu ^* } \right)\psi _{S,R} (x) + \psibar_{S,R} \left( {m_{\text{Sea}}^\dag  - \tfrac{1}
{2}ar^\dag\nabla _\mu  \nabla _\mu ^* } \right)\psi _{S,L}(x), \notag
\end{align}
and
\begin{align}\eqn{SGWchiral}
S_{GW}  = &\, a^4 \sum_x {\psibar _{V,L}\,D_{GW}\,\psi _{V,L}(x)  + \psibar
  _{V,R}\,D_{GW}\,\psi _{V,R} }(x)\notag\\
& \hspace{.5in} + \psibar _{V,L}\,m_{\text{Val}}\,\psi
_{V,R}(x)  + \psibar _{V,R}\,m_{\text{Val}}^\dag\,\psi _{V,L}(x). 
\end{align}
Here, for reasons that will become clearer shortly, we consider $m_\text{Sea}$, $m_\text{Val}$ and $r$ to be matrices in
flavor space, and identify the parameters that appear between 
right-handed anti-quarks and left-handed quarks as their hermitian conjugates.

Clearly, when $m_\text{Sea}=m_\text{Val}=r=0$,
\eqs{SWchiral}{SGWchiral} are invariant under independent global rotations of
the left-handed and right-handed components of all quark fields:
\begin{align}\eqn{FieldTransformation}
\psi _{X,\chi }  \to g_{X,\chi } \psi _{X,\chi } ,\quad \psibar _{X,\chi }  \to \psibar_{X,\chi } g_{X,\chi }^\dag  ,\qquad X = S,V,\quad\chi  = L,R,
\end{align}
where $g_{S,L}$ and $g_{S,R}$ are in $SU(N_f)$, while $g_{V,L}$ and
 $g_{V,R}$ are in $SU(N_V|N_V)$.
 We conclude that flavor transformations
belonging to the group in \eq{Sym_Group_Hybrid} 
are symmetries of the action \eq{SH} 
broken by the mass terms $m_{Val}$ and $m_{Sea}$ and the Wilson term
$r$.

It is convenient to treat these symmetry breaking parameters as spurion fields,
i.e. assuming the transformation rules 
\begin{align} \eqn{SpurionTransformation}
m_{\text{Val}}  & \;\to\; g_{V,L} \,m_{\text{Val}}\, g_{V,R}^\dag\,\nonumber
,\quad m_{\text{Val}}^\dag   \to g_{V,R}\,m_{\text{Val}}^\dag\,g_{V,L}^\dag 
\ ,\\ 
m_{\text{Sea}}  & \;\to\; g_{S,L} \,m_{\text{Sea}}\, g_{S,R}^\dag \,,
\quad m_{\text{Sea}}^\dag   \to g_{S,R}\,m_{\text{Sea}}^\dag\,g_{S,L}^\dag 
 \ ,\\ 
 r &\; \to \;g_{S,L}  \,r \,g_{S,R}^\dag\,,\hspace{.57in} r^\dag\;\to\;\;g_{S,R}
 \,r^\dag \,g_{S,L}^\dag\ .\nonumber 
\end{align}
The mixed action $S_M$, even with non-vanishing mass and Wilson
terms, is invariant
under the combined transformations
\eqs{FieldTransformation}{SpurionTransformation}.

To complete this part we note that for a transformation to be a
symmetry of the theory it must also leave unchanged the integration
measure in the functional integral. It is
a simple matter to show that the measure for the sea Wilson fermions
is invariant under the global rotations in
\eq{FieldTransformation}. 
The situation for the valence quarks is not quite so simple because of
the operator $\hat{\gamma_5}$ in the chiral variation. It turns out, however, as
has been shown in ref. \cite{Luscher}, that the
measure is indeed invariant under the symmetry
transformations considered here -- the flavor non-singlet transformations. 
The last statement can be extended to the full
valence sector, including the
ghost fields.\footnote{The symmetry group that we write here is not
  the true 
  symmetry group of the quantized theory. As discussed in
 \cite{verbaarschot,shshIV}, the 
  presence of ghost fields in the functional integral leads to
 constraints on the allowed 
  symmetry transformations.
  However, for the derivation of the correct chiral
  Lagrangian it is possible to use the symmetry group in
  \eq{Sym_Group_Hybrid} \cite{shshIV}.}

\subsection{Symanzik action}\label{Symanzik}

We construct  Symanzik's local effective theory
which, close to the continuum, describes the same long-range physics as the
discrete lattice action well below the momentum cutoff
$1/a$ \cite{'tHooft:1980xb,Symanzik:1983dc,Symanzik:1983gh}. 

Since the continuum action $S_S$ is designed to reproduce the same
long-range correlation functions as  
the discrete lattice action $S_M$, it must have the same symmetries [\eq{Sym_Group_Hybrid}] as
the underlying theory. 
Up to $\mcal O(a)$, 
the quark operators that enter are of mass dimensions 3,4, and 5,
which include only quark bilinears.
Moreover, the independent symmetry transformations acting separately on the  
sea and valence sectors requires that the quark bilinears do not mix the
sectors.
 This implies that up to $\mcal O(a)$, the fermionic operators in $S_S$
(as in $S_M$) are of two types -- one built of sea quarks only  
and one of valence quarks. 

It is straightforward to write down the $\mcal O(a)$ Symanzik action
$S_S$ using previous results concerning Wilson
fermions~\cite{Sheikholeslami:1985ij,Luscher:1996sc} 
and Ginsparg-Wilson fermions~\cite{Niedermayer:1998bi}.
The
details of the analysis are deferred to appendix~\ref{SS} -- here we
only quote the result (for the fermionic part of the action):
\beq\eqn{LSA}
S_S  = \int d^4x \[ {
 \overline{\psi} _V ( D + \widetilde m_{\text{Val}} )\psi _V
+\overline{\psi} _S ( D + \widetilde m_{\text{Sea}} )\psi _S  
  + ac_{SW} 
\overline{\psi} _S \sigma _{\mu \nu } F_{\mu \nu } \psi _S }\]  + \mathcal{O}(a^2 ).
\eeq
$\widetilde{m}_{\text{Val}}$ and $\widetilde m_{\text{Sea}}$ are 
renormalized masses. Two consequences of the exact chiral
symmetry of the massless Ginsparg-Wilson fermions are: $(a)$ there is no Pauli term
$\psibar\sigma_{\mu \nu} F_{\mu \nu}\psi$ 
for the valence sector, and $(b)$ the valence quark mass is
only multiplicatively renormalized. No symmetry protects the Wilson
sea quarks from getting an additive correction of the order of the
cutoff $1/a$.

It is useful at this point to collect the quark fields in a single
quark field vector $\Psi$, and rewrite~\eq{LSA} as
\beq
& S_S  = \int {\overline{\Psi} ( D +  \mathbf{m})\Psi  + \overline{\Psi} \,a\mathbf{c_{SW}} 
\sigma _{\mu \nu } F_{\mu \nu } \Psi }  + \mathcal{O}(a^2 )
\eqn{LSA2}\ , \\[1ex] 
 & \Psi  = \left( \begin{gathered}
  \psi _S  \hfill \\
  \psi _V  \hfill \\
\end{gathered}  \right),\quad \mathbf{m} = \left( {\begin{matrix}
{\widetilde m_{\text{Sea}} }&{}\\
{}&{\widetilde m_{\text{Val}} }\\
\end{matrix}} \right),\quad a\mathbf{c_{SW}} = \left( {\begin{matrix}
{ac_{SW}}&{}\\
{}&0\\
\end{matrix}} \right).
\eeq

\subsection{Symmetries of $S_S$ and \CPTs for the
mixed action}\label{EFT}

We now turn to the construction of a low-energy effective
theory for the ``underlying'' Symanzik action in \eq{LSA2}. The method
is completely analogous to the construction of the 
chiral Lagrangian in QCD \cite{Weinberg:1979kz}.%
\footnote{It should be noted that as of yet the construction of PQ
  \CPTs from PQ QCD is not as well justified as the standard
  derivation of \CPTs from QCD. The arguments of the latter
  cannot be trivially extended to PQ QCD because they
  rely, in part, on the existence of a Hilbert space of physical
  states with a 
  positive definite norm, which is absent in the presence of ghost
  fields. The same is also true for the mixed
  action. The validity of the \CPTs for the mixed action is thus on
  the same footing
   as that of PQ \CPT, which has been discussed in \cite{shshIV}.}
The idea is that the spontaneous breaking 
of the approximate chiral symmetry
gives rise to light pseudo Goldstone
bosons, the light mesons, which at low energies
are the only
relevant degrees of 
freedom. The effective action is written in terms of local
interactions of the pseudo Goldstone fields. Since the pseudo Goldstone bosons interact weakly at low energies, the action can be organized in a perturbative series.

All observables calculated are expanded in two small parameters,
\beq
\epsilon\sim\frac{p^2}
{\Lambda_\chi^2}\sim\frac{\widehat m}{\Lambda_\chi^2}\quad\mbox{
  and }\quad\delta\sim\frac{\widehat a}{\Lambda_\chi^2}\ ,
\eeq
where 
$p$ is the light meson momentum, 
$\Lambda_\chi\sim 1$GeV is the chiral symmetry breaking scale,
and
$\widehat m$ and $\widehat a$ stand for matrix elements of\footnote{%
Though the notation might obscure this fact, $\mathbf{\widehat m}$ and
$\mathbf{\widehat a}$ both have mass dimension 2 -- 
they are the leading contributions to the squared mass of the pseudo
Goldstone boson. We nevertheless use this notation as it makes the dependence
on the quark masses and the lattice spacing more transparent.
}
\beq
\mathbf{\widehat m}\equiv 2 B_0 \mathbf{m},\quad \mathbf{\widehat a}\equiv2 W_0 a\mathbf{\CSW}.
\eeq
$B_0$ and $W_0$ are dimensionful 
low-energy constants that appear in the effective theory at leading
order [see \eq{L2}]. 
  They depend only on the high-energy scale
$\Lambda_\chi$, and dimensional analysis reveals that the perturbative
expansion is in fact in 
$m_q/\Lambda_\chi$ and $a\Lambda_\chi$. 
We follow ref.~\cite{Rupak:2002sm} and choose 
the $\{\epsilon, \delta\}$ terms as leading order (LO) and the 
$\{\epsilon^2,\epsilon\delta\}$ terms as NLO in the effective Lagrangian, 
dropping $\mcal O(a^2)$ contributions. 
The underlying hierarchy consistent with this ordering is
$\left\{ {\epsilon ,\delta } \right\} \gg \left\{ {\epsilon ^2 ,\epsilon \delta } \right\} \gg \delta ^2$, 
and the last inequality also implies $\epsilon\gg\delta$. 
This ordering is chosen for convenience and is somewhat arbitrary. In
practice, the double expansion should be organized according to the
actual relative sizes of the quark masses and the lattice spacing.

 The effective Lagrangian is constructed from all operators that 
 respect the symmetries of the underlying action
     $S_S$.
The compact notation of \eq{LSA2} makes it easy to see that the
symmetry group of $S_S$ to $\mcal{O}(a)$ is
\begin{align}\eqn{Sym_WCPT}
SU(N_f  + N_V |N_V )_L  \otimes SU(N_f  + N_V |N_V )_R.
\end{align} 

This symmetry group 
(treating $\mathbf{\widehat m}$ and $\mathbf{\widehat a}$ as spurion fields) 
is the same as that of PQ
QCD with Wilson fermions. Indeed the effective action 
$S_S$ in~\eq{LSA2} is the same as the effective Symanzik action
for the PQ QCD
Wilson action of ref.~\cite{Rupak:2002sm}, with a specific choice of
$a\mathbf{\CSW}$ that has support only in the sea sector.
This fortunate similarity between the mixed action
theory and the PQ QCD Wilson theory implies that
\emph{ the mixed low-energy
chiral effective action
  has the same structure, at
  $\mcal{O}(a)$, 
as the action of Wilson \CPT, introduced in ref.~\cite{Rupak:2002sm}, with the
  restriction that $a\mathbf{\CSW}$
vanishes in the valence-ghost sector.}
The chiral
Lagrangian to $\mcal{O}(a)$, which describes both W\CPTs and the mixed
effective theory, is provided in appendix~\ref{WChPT}.

\subsection{Application: Meson mass}
\label{applications}

In this subsection we give an example for the use of the
mixed chiral Lagrangian. For simplicity we take all the sea quarks
and all the valence quarks to be (separately) 
degenerate, and the Wilson parameter to be
a flavor singlet in the sea sector. This amounts to setting
\begin{align}
\eqn{degvalues}
\mathbf{\widehat m} = \operatorname{diag} \left({\widehat m_{\text{Sea}} ,\widehat m_{\text{Val}}}\right),\quad
\mathbf{\widehat a} = \operatorname{diag} \left({\widehat a,0}\right).
\end{align}
The number of sea
quark flavors is taken to be $N_f=3$.
We consider the expression for the mass of the flavor charged meson
with valence quark flavor indices $AB$ ($A\ne B$) to NLO. Using the
relation between the mixed chiral effective theory and 
W\CPT, one can obtain the result straightforwardly by taking
the mass formula from ref.~\cite{Rupak:2002sm} with the values for
$\mathbf{\widehat m}$ and $\mathbf{\widehat a}$ 
given by \eq{degvalues}. We find
\begin{align}
\eqn{Hdegmass}
\left( {M_{\text{AB}}^\text{2} } \right)_{\text{NLO}}  &=   \widehat m_{\text{Val}}  + \frac{{\widehat m_{\text{Val}} }}
{{48f^2 \pi ^2 }}\left[ {\widehat m_{\text{Val}}  - \widehat m_{\text{Sea}}  - \widehat a + (2\widehat m_{\text{Val}}  - \widehat m_{\text{Sea}}  - \widehat a)\ln \left( {\widehat m_{\text{Val}} } \right)} \right] \\ 
   & \hspace{1in} - \frac{{8\widehat m_{\text{Val}} }}
{{f^2 }}\left[ {(L_5  - 2L_8 )\widehat m_{\text{Val}}  + 3(L_4  - 2L_6 )\widehat m_{\text{Sea}}  + 3(W_4  - W_6 )\widehat a} \right].\notag
\end{align}
Here, the parameters $L_i$ are the Gasser-Leutwyler coefficients, and
$W_4$ and $W_6$ are additional low-energy constants that enter the
chiral Lagrangian at NLO.
Note that for $\widehat a=0$ the expression for PQ \CPTs
(calculated in \cite{shPQ,GLPQ,Sharpe:2000bc})
is recovered. 

\Eq{Hdegmass} demonstrates the analytic connection between QCD and the
simulated mixed action theory. It shows the latter to be a calculation with
controlled systematic errors. From fitting the equation to the
appropriate data from numerical simulations one can obtain an estimate
for the linear combinations of Gasser-Leutwyler coefficients that appear in it. 

Examining \eq{Hdegmass} one can also appreciate the
potential advantage of 
using a mixed lattice action. 
In simulations using Wilson fermions in both sea and valence sectors,
an equation similar to \eq{Hdegmass} holds (see \eq{Wdegmass}).
In that case, the range of valence quark masses that can be
simulated  might be too small to convincingly show the curvature coming
from the quadratic dependence and the logarithms that enter at NLO. 
By using Ginsparg-Wilson fermions for the valence quarks one can vary the valence
quark masses over a wider range. The expected NLO curvature, on which
the extraction of 
the Gasser-Leutwyler coefficients depends, is consequently much more
likely to be 
seen. 

Finally, comparison with the result for W\CPT,
\eq{Wdegmass}, reveals that the latter depends on twice the number of
$W_i$ coefficients. This is fortunate for the mixed theory as it
makes the predictions of 
the effective theory less dependent on parameters that have no
particular relevance to QCD.

To understand this simplification in the expression for the meson
mass, consider the relation between
the symmetries of the
mixed action and those of PQ Wilson action. On the one hand the
massless mixed theory has exact chiral symmetry in the 
valence sector, which the Wilson action does not. On the other hand,
the valence and sea quarks of the Wilson action have the same type of
Dirac operator which allows mixing between the sectors - a
transformation which is not a symmetry of the mixed action formulation.
At $\mcal{O}(a)$, however, the breaking of the sea-valence symmetry in the
mixed theory does not yet show up, and thus the simpler expressions
arise due to the larger chiral symmetry.

\end{section}
\begin{section}{Summary}
\label{summary}

In this paper we discuss lattice simulations with different fermions
for sea and valence quarks.  
As a particular example we have studied here the case with Wilson sea
quarks and Ginsparg-Wilson valence quarks. Using Symanzik's effective
action for lattice theories as an intermediate step, we have
derived the form of the low-energy
chiral Lagrangian for the mixed theory to $\mcal{O}(a)$. The construction
shows that simulations with the
mixed action provide as controlled an approximation to QCD as
partially quenched simulations. This is to be expected since the mixed action reduces to PQ QCD in the continuum limit. 

The goal of the mixed action approach is similar to that of PQ
QCD. The use of smaller valence quark masses allows one to 
probe deeper the chiral regime of QCD 
and obtain additional information on the low-energy constants, the
Gasser--Leutwyler coefficients. 
Furthermore, the use of chiral lattice fermions in the valence sector,
instead of Wilson fermions,
makes it possible to simulate much lighter valence quarks.
This leads to more data points obtained on the lattice and consequently
to more reliable fits of \CPTs to the data.

Here we have demonstrated the mixed action approach for Wilson sea
quarks and Ginsparg-Wilson valence quarks taking into account the
leading $\mcal O(a)$ contributions.  
An important extension of this analysis is the inclusion of $\mcal
O(a^2)$ effects.
First of all, the lattice spacing determined by the unquenched configurations is possibly not small enough to safely neglect the $\mcal O(a^2)$ corrections. If one wants to fit the lattice data directly to equations like \eq{Hdegmass} without taking the continuum limit first, the $\mcal O(a^2)$ corrections should be included to obtain better fits.
Second, the $\mcal{O}(a)$ effects are generated here only by the
Wilson sea quarks. Many unquenched simulations are in fact performed with
non-perturbatively $\mcal{O}(a)$-improved Wilson fermions for the sea
quarks. The leading corrections for these simulations are of $\mcal
O(a^2)$ and need to be computed in order to know how the continuum
limit is approached.   

While valuable, it should also be noted that the inclusion of $\mcal
O(a^2)$ effects in the chiral Lagrangian framework is likely to be a
hard task. The main difficulty arises from the many new operators that
enter the Symanzik action at this order. Some of these operators break
Lorentz invariance, while several others break the chiral symmetry and
require the introduction of additional spurion fields.

The approach proposed here should be also studied with other
combinations for the lattice fermions. In particular, the case with
staggered sea quarks is interesting, since staggered fermions are
computationally cheaper. 
At present the lightest dynamical
quark masses are achieved with staggered fermions.
It is well-known that applying staggered fermions to QCD involves a
theoretical uncertainty and is possibly uncontrolled. Consequently, predictions from
chiral perturbation theory for staggered fermions would also serve as a
test of this discretization method \cite{BGPQ}.

Finally, the cost of simulations of a mixed action is roughly the sum of the cost of
generating a set of unquenched gauge field configurations plus that of
analyzing quenched
simulations with Ginsparg-Wilson fermions. Thus, we can expect that in
the near future 
simulations with a mixed action
will become feasible.

\end{section}
\begin{acknowledgments}

We acknowledge support in part by U.S. DOE grants
DF-FC02-94ER40818, DE-AC03-76SF00098, 
DE-FG03-96ER40956/A006 and DE-FG02-91ER40676. 
We thank Maarten Golterman and Steve Sharpe for their comments on this manuscript.
G.R. would like to thank the
Department of Physics, Boston University and the 
Benasque Center for Science, Benasque, Spain for kind hospitality during part
of this work. 

\end{acknowledgments}
\appendix

\begin{section}{Symanzik action for the Wilson and Ginsparg-Wilson actions}
\label{SS}
 
In this appendix we derive \eq{LSA} for
the Symanzik action describing the mixed theory to $\mcal{O}(a)$.
As has been stated in the text, to this order the Symanzik action is 
simply the sum of the local effective actions for the valence and
the sea sectors.
 
The local Symanzik action for Wilson fermions has been derived
in~\cite{Sheikholeslami:1985ij,Luscher:1996sc}.  
One first lists all the operators of mass
dimension
no greater than 5, which respect the symmetries of the Wilson lattice
action 
(the appropriate power of $a$ is inserted to complete the dimensions
of terms in the Lagrangian to 4).
The operators of dimension 4 (which are $a$ independent) make up, by
construction, the 
continuum action of QCD. 
 
Because the Wilson term explicitly breaks the chiral symmetry, it is
expected that the quark mass be additively renormalized, and the
size of the correction should be of the order of the cutoff scale $1/a$.
Indeed, the
only dimension 3 operator is 
$\psibar\psi$, which appears in the action with a coefficient
proportional to $1/a$ and has precisely this effect.
 
There are several operators of mass dimension 5 that are allowed by
the symmetries. Some of these operators can be eliminated using the
leading
order equations of motion. Others have the same structure as the mass
and kinetic operators
that already appear in the QCD action, and have the effect of
renormalizing the quark masses
and the gauge coupling.
Finally, a single term is left -- the Pauli
term: $\psibar\sigma_{\mu\nu}F_{\mu\nu}\psi$. Note that the Pauli
term breaks the chiral symmetry, and is therefore allowed only because
of the Wilson term. Putting it all together, the Symanzik action
for the Wilson sea sector is
\beq\eqn{LSAWilson}
\int d^4x \[\psibar _S (D + \widetilde m_{\text{Sea}} )\psi _S  
  + ac_{SW} 
\psibar _S \sigma _{\mu \nu } F_{\mu \nu } \psi _S\]  +
\mathcal{O}(a^2 )\ .
\eeq
where $\widetilde m_{\text{Sea}}$ is the renormalized sea quark mass, and
$c_{SW}$ is an unknown coefficient.
 
The analysis for the Ginsparg-Wilson valence quarks is similar. This
may seem confusing due to the fact that some of the chiral 
projectors on the lattice are written in terms of $\hat{\gamma_5}$,
and not $\gamma_5$ as in the continuum theory. However, it has been
shown in ref. \cite{Hasenfratz1, Hasenfratz2},
that the chiral symmetry of the Ginsparg-Wilson lattice
action leads to exactly the same chiral Ward identities which appear in the
continuum. Hence, by imposing the usual chiral symmetry on the Symanzik
action, the effective theory correctly reproduces the consequences of
the lattice chiral symmetry.

Due to the exact chiral symmetry the valence quark
mass gets renormalized only multiplicatively and
the Pauli term is absent. Consequently, after considering the
renormalizations of gauge coupling and quark masses, the Symanzik
action for the valence Ginsparg-Wilson quarks contains no $\mcal{O}(a)$ part (the Ginsparg-Wilson
lattice action is automatically $\mcal{O}(a)$
improved\cite{Niedermayer:1998bi}):
\beq\eqn{LSAGW}
\int d^4x \[\psibar _V (D + \widetilde m_{\text{Val}} )\psi _V\] +
\mathcal{O}(a^2 )\ .
\eeq
 
\Eq{LSA} is the sum of \eq{LSAWilson} and \eq{LSAGW}.
 
\end{section}
\begin{section}{W\CPTs results}
\label{WChPT}
 
We present the W\CPTs Lagrangian which also
 describes the mixed theory to $\mcal O(a)$. Interested readers should
consult ref.~\cite{Rupak:2002sm} for further details on W\CPT. 
We also provide the expression for 
the mass of a flavor charged meson for
 comparison with the mixed theory result. 
  
The W\CPTs Lagrangian is constructed out of operators that
respect
all the symmetries of the underlying theory 
in~\eq{LSA2}, with explicit flavor axial symmetry breaking terms
constructed out 
of $\widehat m$ and ${\widehat a}$. 
As described in the text, the LO Lagrangian is linear in $\epsilon$ and
$\delta$:  
\beq\eqn{L2}
\mathcal{L}_2  &=& \frac{{f^2 }}
{4}\left\langle \partial \Sigma\partial \Sigma^\dag  \right\rangle -
\frac{{f^2 }}
{4}\left\langle(\widehat m  + {\widehat a} )\Sigma^\dag   
+ \Sigma(\widehat m ^\dag   + {\widehat a} ^\dag 
 )\right\rangle\ .
\eeq
Here the angled brackets stand for the super-trace over the flavor
indices:
\begin{align}
\left\langle \Gamma  \right\rangle  = \operatorname{str} \left( \Gamma
\right) = \sum\limits_i {\eta _i \Gamma _{ii} } ,\quad \eta _i  =
\left\{ {\begin{matrix}
1&{i\text{ is a quark flavor index}}\\
{\text{ - 1}}&{i\text{ is a ghost flavor index}}\\
\end{matrix}} \right.\,\,\,\,.
\end{align}
and $\Sigma=\exp\(2 i\Pi/f\)$ is a non-linear representation of the
meson
fields.
 
The NLO
Lagrangian is:
\beq\eqn{L4} 
\mcal L_4&=&
{L_1 \left\langle {\partial \Sigma\partial \Sigma^\dag  } \right\rangle
^2 }+
{L_2 \left\langle {\partial _\mu  \Sigma\partial _\nu  \Sigma^\dag  }
  \right\rangle 
 \left\langle {\partial _\mu  \Sigma\partial _\nu  \Sigma^\dag  }
 \right\rangle }+ 
{L_3 \left\langle {(\partial \Sigma\partial \Sigma^\dag  )^2 }
  \right\rangle }\nonumber\\ 
&{}&+{L_4 \left\langle {\partial \Sigma\partial \Sigma^\dag  }
 \right\rangle \left\langle 
{\widehat m^\dag  \Sigma + \Sigma^\dag  \widehat m } \right\rangle }
+{W_4 \left\langle {\partial \Sigma\partial \Sigma^\dag  }
  \right\rangle \left\langle  
{{\widehat a} ^\dag  \Sigma + \Sigma^\dag  {\widehat a} } \right\rangle
}\nonumber\\
&{}&+{L_5 \left\langle {\partial \Sigma\partial \Sigma^\dag 
 (\widehat m ^\dag  \Sigma + \Sigma^\dag  \widehat m )} \right\rangle }
+{W_5 \left\langle {\partial \Sigma\partial \Sigma^\dag  ({\widehat a} ^\dag
\Sigma + \Sigma^\dag  {\widehat a} )} \right\rangle }\nonumber\\
&{}&+{L_6 \left\langle {\widehat m ^\dag  \Sigma + \Sigma^\dag  \widehat m }
  \right\rangle ^2 } 
+{W_{6} \left\langle {\widehat m ^\dag  \Sigma + \Sigma^\dag  \widehat m }
  \right\rangle \left\langle {{\widehat a} ^\dag  \Sigma + \Sigma^\dag
{\widehat a} }
  \right\rangle }\nonumber\\ 
&{}&+ {L_7 \left\langle {\widehat m ^\dag  \Sigma - \Sigma^\dag  \widehat m }
  \right\rangle ^2 } 
+{W_{7} \left\langle {\widehat m ^\dag  \Sigma - \Sigma^\dag  \widehat m }
\right\rangle 
\left\langle {{\widehat a} ^\dag  \Sigma - \Sigma^\dag  {\widehat a} }
\right\rangle
}\nonumber\\ 
&{}&+{L_8 \left\langle {\widehat m ^\dag  \Sigma\widehat m ^\dag  \Sigma +
\Sigma^\dag  
\widehat m \Sigma^\dag  \widehat m } \right\rangle }
+{W_{8} \left\langle {{\widehat a} ^\dag  \Sigma\widehat m ^\dag  \Sigma +
\Sigma^\dag 
 {\widehat a} \Sigma^\dag  \widehat m } \right\rangle }\ .
\eeq
These
Lagrangians describe both the mixed and the PQ Wilson lattice
actions.  
In the mixed theory ${\widehat a}$ has
support only in the sea-sea sector. We comment that \eqsii{L2}{L4}
contains
ordinary \CPT. Moreover, since the low-energy constants $L_i$'s and
$W_i$'s are  
independent  of $\widehat m$ and $\widehat a$ and this theory becomes 
the familiar
\CPTs in the sea-sea
sector when $a\rightarrow 0$, the $L_i$'s are exactly the
Gasser-Leutwyler
 coefficiants of ordinary \CPTs.  
 
Next, we provide the  W\CPT~expression for the 
 mass of the flavor 
charged meson defined in Secion~\ref{applications}. We consider the
 case where
\begin{align}
\mathbf{\widehat m} = \operatorname{diag} \left( {\widehat m_{\text{Sea}}
   ,\widehat m_{\text{Val}} } \right),\quad
\mathbf{\widehat a} = \operatorname{diag} \left( {\widehat a_{\text{Sea}}
   ,\widehat a_{\text{Val}} } \right)
\end{align}
(compare with \eq{degvalues}).
One obtains:
\begin{align}
\eqn{Wdegmass}
\left( {M_{\text{AB}}^\text{2} } \right)_{\text{NLO}}  =  & (\widehat m_{\text{Val}}  + \widehat a_{\text{Val}} ) \notag 
     + \frac{{(\widehat m_{\text{Val}}  + \widehat a_{\text{Val}} )}}
{{48f^2 \pi ^2 }}\left[ {(\widehat m_{\text{Val}}  + \widehat a_{\text{Val}} ) - (\widehat m_{\text{Sea}}  + \widehat a_{\text{Sea}} )} \right.\notag  \\ 
   & \hspace{1.5in}\left. { + \left( {2(\widehat m_{\text{Val}}  + \widehat a_{\text{Val}} ) - (\widehat m_{\text{Sea}}  + \widehat a_{\text{Sea}} )} \right)\ln \left( {\widehat m_{\text{Val}}  + \widehat a_{\text{Val}} } \right)} \right]\notag  \\ 
   &  - \frac{{8\widehat m_{\text{Val}} }}
{{f^2 }}\left[ {(L_5  - 2L_8 )\widehat m_{\text{Val}}  + 3(L_4  - 2L_6 )\widehat m_{\text{Sea}}  + 3(W_4  - W_6 )\widehat a_{\text{Sea}} } \right]\notag  \\ 
   &  - \frac{{8\widehat a_{\text{Val}} }}
{{f^2 }}\left[ {(L_5  + W_5  - 2W_8 )\widehat m_{\text{Val}}  + 3(L_4  - W_6 )\widehat m_{\text{Sea}} } \right]
\end{align}
To obtain the expression appropriate for common lattice simulations, in
which the Wilson term is the same for all flavors, one sets $\widehat
a_\text{Val}=\widehat a_\text{Sea}$ in the last equation. The meson mass
for the mixed action [\eq{Hdegmass}] can be obtained by setting $\widehat a_\text{Val}=0$.

\end{section}


\begin{thebibliography}{10}

\bibitem{Gasser:1985gg}
J.~Gasser and H.~Leutwyler,
\newblock Nucl. Phys. {\bf B250}, 465 (1985).

\bibitem{Sharpe:2000bc}
S.~R. Sharpe and N.~Shoresh,
\newblock Phys. Rev. {\bf D62}, 094503 (2000), [hep-lat/0006017].

\bibitem{Cohen:1999kk}
A.~G. Cohen, D.~B. Kaplan and A.~E. Nelson,
\newblock JHEP {\bf 11}, 027 (1999), [hep-lat/9909091].

\bibitem{Bernard:2002yk}
C.~Bernard {\em et~al.},
\newblock hep-lat/0209086.

\bibitem{Durr:2002fe}
S.~D{\"u}rr,
\newblock hep-ph/0209319.

\bibitem{Farchioni:2002ca}
qq+q, F.~Farchioni, C.~Gebert, I.~Montvay and L.~Scorzato,
\newblock hep-lat/0209142.

\bibitem{BGPQ}
C.~W. Bernard and M.~F.~L. Golterman,
\newblock Phys. Rev. {\bf D49}, 486 (1994), [hep-lat/9306005].

\bibitem{Sharpe:1998xm}
S.~R. Sharpe and J.~Singleton, Robert,
\newblock Phys. Rev. {\bf D58}, 074501 (1998), [hep-lat/9804028].

\bibitem{Rupak:2002sm}
G.~Rupak and N.~Shoresh,
\newblock Phys. Rev. {\bf D66}, 054503 (2002), [hep-lat/0201019].

\bibitem{Lee:1999zx}
W.-J. Lee and S.~R. Sharpe,
\newblock Phys. Rev. {\bf D60}, 114503 (1999), [hep-lat/9905023].

\bibitem{Aubin:2002ss}
C.~Aubin {\em et~al.},
\newblock hep-lat/0209066.

\bibitem{Wilson:1974sk}
K.~G. Wilson,
\newblock Phys. Rev. {\bf D10}, 2445 (1974).

\bibitem{Ginsparg:1982bj}
P.~H. Ginsparg and K.~G. Wilson,
\newblock Phys. Rev. {\bf D25}, 2649 (1982).

\bibitem{Hasenfratz:1998ri}
P.~Hasenfratz, V.~Laliena and F.~Niedermayer,
\newblock Phys. Lett. {\bf B427}, 125 (1998), [hep-lat/9801021].

\bibitem{Neuberger:1998fp}
H.~Neuberger,
\newblock Phys. Lett. {\bf B417}, 141 (1998), [hep-lat/9707022].

\bibitem{Neuberger:1998wv}
H.~Neuberger,
\newblock Phys. Lett. {\bf B427}, 353 (1998), [hep-lat/9801031].

\bibitem{Niedermayer:1998bi}
F.~Niedermayer,
\newblock Nucl. Phys. Proc. Suppl. {\bf 73}, 105 (1999), [hep-lat/9810026].

\bibitem{Luscher}
M.~Luscher,
\newblock Phys.\ Lett.\ B {\bf 428}, 342 (1998), [hep-lat/9802011].

\bibitem{verbaarschot}
P.~H. Damgaard, J.~C. Osborn, D.~Toublan and J.~J.~M. Verbaarschot,
\newblock Nucl. Phys. {\bf B547}, 305 (1999), [hep-th/9811212].

\bibitem{shshIV}
S.~R. Sharpe and N.~Shoresh,
\newblock Phys. Rev. {\bf D64} (2001), [hep-lat/0108003].

\bibitem{'tHooft:1980xb}
K.~Symanzik,
\newblock New York, Usa: Plenum (1980) 438 P. (Nato Advanced Study Institutes
  Series: Series B, Physics, 59).

\bibitem{Symanzik:1983dc}
K.~Symanzik,
\newblock Nucl. Phys. {\bf B226}, 187 (1983).

\bibitem{Symanzik:1983gh}
K.~Symanzik,
\newblock Nucl. Phys. {\bf B226}, 205 (1983).

\bibitem{Hernandez:1998et}
P.~Hern\'{a}ndez, K.~Jansen and M.~L{\"u}scher,
\newblock Nucl. Phys. {\bf B552}, 363 (1999), [hep-lat/9808010].

\bibitem{Sheikholeslami:1985ij}
B.~Sheikholeslami and R.~Wohlert,
\newblock Nucl. Phys. {\bf B259}, 572 (1985).

\bibitem{Luscher:1996sc}
M.~L{\"u}scher, S.~Sint, R.~Sommer and P.~Weisz,
\newblock Nucl. Phys. {\bf B478}, 365 (1996), [hep-lat/9605038].

\bibitem{Weinberg:1979kz}
S.~Weinberg,
\newblock Physica {\bf A96}, 327 (1979).

\bibitem{shPQ}
S.~R. Sharpe,
\newblock Phys. Rev. {\bf D56}, 7052 (1997), [hep-lat/9707018],
\newblock Erratum-ibid.D62:099901,2000.

\bibitem{GLPQ}
M.~F.~L. Golterman and K.-C. Leung,
\newblock Phys. Rev. {\bf D57}, 5703 (1998), [hep-lat/9711033].

\bibitem{Hasenfratz1}
P.~Hasenfratz,
\newblock Nucl.\ Phys.\ B {\bf 525}, 401 (1998), [hep-lat/9802007].

\bibitem{Hasenfratz2}
P.~Hasenfratz, S.~Hauswirth, T.~Jorg, F.~Niedermayer and K.~Holland,
Nucl.\ Phys.\ B {\bf 643}, 280 (2002), [hep-lat/0205010].

\end{thebibliography}
\end{document}